\documentclass[aps,prd,amsmath,amssymb,preprintnumbers,preprint,nofootinbib,a4paper]{revtex4-1}
\pdfoutput=1
\usepackage{amsthm}
\usepackage{graphicx}
\usepackage{color}
\newcommand{\beq}{\begin{eqnarray}}
\newcommand{\eeq}{\end{eqnarray}}

\newcommand{\bmp}{\noindent\begin{minipage}{16cm}}
\newcommand{\emp}{\end{minipage}\vskip 7mm} 

\usepackage{dcolumn}
\usepackage{bm}
\usepackage{bbm}
\usepackage{subfigure}
\usepackage{pxfonts}
\usepackage{epstopdf}

\theoremstyle{definition}

\theoremstyle{plain}

\usepackage{epsfig}

\usepackage[ margin=5pt, font=normalsize,labelfont=bf,justification=raggedright]{caption}
\usepackage{youngtab}
\usepackage{slashed}
\def\lsim{\mathrel{\rlap{\lower4pt\hbox{\hskip1pt$\sim$}}
    \raise1pt\hbox{$<$}}}                
\def\gsim{\mathrel{\rlap{\lower4pt\hbox{\hskip1pt$\sim$}}
    \raise1pt\hbox{$>$}}}                

\def\etm{ E_T \! \! \! \! \! \! / \, \,  \,}
\def\ga{\gamma}
\def\gev{\textrm{ GeV }}

\preprint{CERN_PH_TH-2012-117\\
LAPTH-021/12}

        \def\beq{\begin{equation}}
\def\eeq{\end{equation}}
\def\bea{\begin{eqnarray}}
\def\eea{\end{eqnarray}}
\def\bq{\begin{quote}}
\def\eq{\end{quote}}
\def\to{\rightarrow}
\def\sq{\tilde{q}}
\begin{document}
\preprint{CERN-PH-TH-2012-117}
\preprint{LAPTH-021/12}

\vspace*{1cm}
\title{Model-Independent Bounds on Squarks from Monophoton Searches}
\author{Genevi\`eve B\'elanger$^a$, Matti Heikinheimo$^{a,b,c}$ and Ver\'onica Sanz$^{b,c}$}
\affiliation{$^a$ LAPTH, U. de Savoie, CNRS, BP 110, 74941 Annecy-Le-Vieux, France \\
$^b$ Theory Division, Physics Department, CERN, CH-1211 Geneva 23, Switzerland \\
$^c$ Department of Physics and Astronomy, York University, Toronto, ON, Canada\\
}
\begin{abstract}
Supersymmetry with a compressed spectrum could be responsible for the negative results from supersymmetric searches at LHC. Squarks and gluinos well below the TeV scale could have escaped detection since all search channels lose sensitivity when the mass splitting between supersymmetric particles becomes small. Even in this stealthy situation, production of colored particles is probed in processes with supersymmetric particles produced in association with a high-pT photon. We show that searches for missing energy with a monophoton are a powerful tool, and that the 2011 LHC data already surpasses the limits set by LEP and TeVatron in the compressed case. We set a model-independent bound on the mass of any up (down) type squark of 150 (110) GeV, and stronger model-dependent bounds can be set. We also comment on the expected improvement on those bounds in the 2012 LHC run.  
\end{abstract}
\maketitle

\section{Introduction}

Supersymmetric models are well motivated extensions of the standard model   both at the theoretical and phenomenological level. 
Not only can supersymmetry stabilize the scalar sector but also models with softly broken supersymmetry can explain unification of gauge couplings and electroweak symmetry breaking, predict a Higgs scalar in the mass range that is compatible with the hints observed at the LHC~\cite{ATLAS:2012ae,Chatrchyan:2012tx} and provide  a good dark matter candidate, the neutral lightest supersymmetric particle (LSP). 
Insisting on having a solution to the hierarchy problem imposes that supersymmetric particles are around the TeV scale. Thus searches for superpartners have taken a prominent part in the physics program of large colliders from LEP, Tevatron to the LHC. As of now there is  no evidence for supersymmetric particles either through direct searches or through their indirect  effect in the flavour sector~\cite{LHCb}.
The lower limits on coloured particles are particularly impressive, indeed squarks and gluinos at the TeV scale should be produced copiously at hadron colliders. Searches using final states containing jets and missing transverse momentum  provide the most powerful limits. In the context of the constrained minimal supersymmetric standard model (CMSSM) 
with an integrated luminosity of $4.7 fb^{-1}$ the  95\% CL exclusions for squarks and gluinos of equal mass are 1.4TeV (ATLAS)~\cite{ATLAS_CMSSM} and 1.35GeV~\cite{CMS_CMSSM}.  
 In simplified models  which assume that all squarks of the first two generations are degenerate and decay 100\% in a quark plus missing energy, the same search channel lead to a 95\% CL exclusion for squarks of 1380 GeV  and for gluinos of 940 GeV ~\cite{ATLAS_CMSSM}.

Specific searches for sbottom and stop squarks were also conducted. While the limits are not nearly as stringent, these searches are important because radiative corrections from the third generation contribute significantly to the  light Higgs mass. 
The ATLAS collaboration performed a search for  sbottom squarks, assuming that the sbottom decays exclusively into a b-quark and a stable neutralino, the 95\% C.L. upper limit 
is $m_{\tilde b_1}>$ 390 GeV for neutralino masses below 60 GeV with ${\cal L}=2.05 {\rm fb}^{-1}$~\cite{Aad:2011cw}.
Similar searches were performed at the Tevatron, the limit set by CDF with ${\cal L}=2.65 {\rm fb}^{-1}$ is 
230 GeV for neutralino masses below 70 GeV~\cite{Aaltonen:2010dy} while D0 with ${\cal L}=5.2 {\rm fb}^-1$ sets a limit of $m_{\tilde b_1} <$ 247 GeV for a massless neutralino~\cite{Abazov:2010wq}. 
The most stringent limits on the stop are obtained in searches from stop produced in gluino decays, such limits are however dependent on the gluino mass~\cite{ATLAS_stop}. 
Direct searches of pair produced stops were performed at the Tevatron.
The channel where the stop decays into charm and neutralino  leads to $m_{\tilde t_1}>180 {\rm GeV}$ for $m_{\tilde \chi_0}=90 {\rm GeV}$  (from CDF with ${\cal L}=2.6 {\rm fb}^{-1}$)~\cite{Aaltonen:2012tq} while the search for $\tilde{t}_1\rightarrow bl\tilde\nu$ excludes  $m_{\tilde t_1}<210 {\rm GeV}$ for a 110 GeV sneutrino (from D0 with  ${\cal L}=5.4 {\rm fb}^{-1}$)~\cite{Abazov:2012cz}.

Despite these impressive exclusion limits, squarks and gluinos well below the TeV scale  could have escaped detection. Indeed all search channels loose sensitivity when the mass splitting between supersymmetric particles becomes small. In this case, the jets and/or leptons that are produced in the supersymmetric decay chains can be very soft and therefore do not pass the basic cuts. 
For example the search for sbottoms with missing transverse momentum and two jets  at hadron colliders that was mentioned above leads to much weaker limits for a small mass splitting between the sbottom and the neutralino.  Full exclusion limits are presented in the plane $m_{\tilde b_1}- m_{\tilde \chi_0}$~\cite{Aad:2011cw}.  To give an idea of the mass splitting probed, let us mention that  CDF and D0 exclude  100 GeV squarks  only when the neutralino LSP mass is below roughly 70 GeV, that D0 probes $160< m_{\tilde b_1}<200$ GeV only if $m_{\tilde \chi_0}<110 {\rm GeV}$~\cite{Abazov:2010wq}
 and  that ATLAS excludes  the range $275 < m_{\tilde b_1}< 350$ ~GeV only for $m_{\tilde \chi_0}<100$~GeV ~\cite{Aad:2011cw}.
 
Experiments are working on closing those cracks by extending their analysis to regions of small mass splitting, for example 
the soft lepton analysis by ATLAS~\cite{softlepton}. 
Another possibility is to use initial state radiation of  a jet or a photon. 
For example the  channel where a jet from the initial state as well as  b-quark jets are tagged has been suggested to improve the LHC sensitivity to mass splitting of  b-squarks and neutralinos of 10 GeV~\cite{Alvarez:2012wf}. 
Nevertheless scenarios where squarks are nearly degenerate with the neutralino LSP remain hard to probe, 
even LEP exclusions are lifted when $m_{\tilde q}-m_{\tilde \chi_0}<m_q$~\cite{LEP_combined}. The purpose of this paper is to set a limit on squarks using the missing energy and a hard photon. This limit is model independent, as we discuss in the next sections.


This paper is organised as follows, section II reviews the compressed spectra scenarios and their experimental and dark matter limits. Section III presents the CMS analysis on monophotons and missing energy. The results of our simulation for one or more degenerate squarks are presented in section IV. The model independent bounds on nearly degenerate squark masses thus obtained are then compared with other search channels  in the conclusion. 

\section{Searches for Compressed Spectra}

Compressed supersymmetric spectra~\cite{Martin:2007gf}  occur in models with non-universal gaugino masses.
Non-universality conditions  at the GUT scale can be found in  supergravity models or supersting-inspired models  ~\cite{Chamoun:2001in,Brignole:1995fb}.  
In models with mixed moduli-anomaly mediated SUSY breaking, gaugino mass splitting is introduced  from anomaly mediation unification ~\cite{Choi:2005uz,Falkowski:2005ck}. 
In non-uninversal models which lead to $M_3/M_2<3$ at the electroweak scale,  all  coloured particles  have a mass that is much closer to that of the sleptons and electroweak-inos  than in the conventional CMSSM. Furthermore the lightest stop is typically the lightest squark and can be nearly degenerate with the neutralino LSP. 
Generically, compressed spectra are also found in  the pMSSM with 19 free parameters defined at the weak scale. Since the soft terms are set arbitrarily, any of the squark can be light. Scenarios with a small mass difference between a squark and the neutralino LSP can be encountered in global scans of the pMSSM after imposing various collider and dark matter constraints. 
In nearly degenerate squark-neutralino LSP scenarios,  the upper limit on the relic density can be easily satisfied.    Indeed when the mass splitting is around 15-20\% (check number) co-annihilation processes involving squarks ( $\tilde{q}\tilde{q} {\rm or}
\tilde{q}\tilde{\chi}$) contribute significantly to reduce the value of the dark matter relic density below the WMAP bound, $\Omega h^2=0.1123$~\cite{Komatsu:2010fb}. 

 This therefore raises the question whether light squarks could still be there and would have been missed. 
Our objective is to look for a model independent bound on squarks. For this we make use of the photon radiated from any leg in the pair production of squarks from gluon or quark pairs.  
 To present our results in a way as model independent as possible, we adopt a simplified model with a single squark and neutralino. The neutralino is kept near-degenerate with  the squark, and the degeneracy is parametrized by
\bea
\Delta m = m_{\sq} - m_{\tilde{\chi}^0} \ .
\eea

\begin{figure}[h!]
\centering
\includegraphics[scale=0.2]{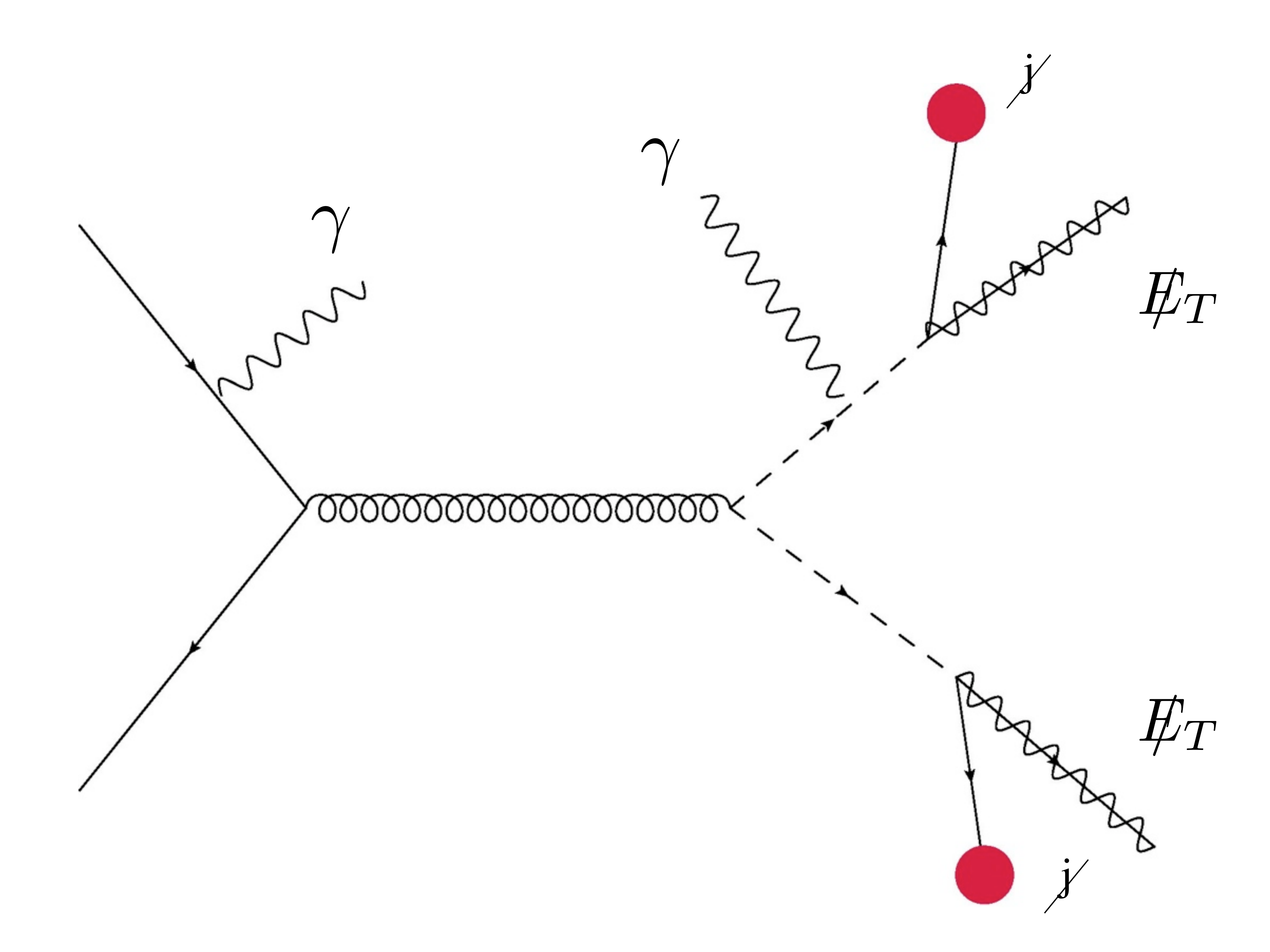}
\includegraphics[scale=0.2]{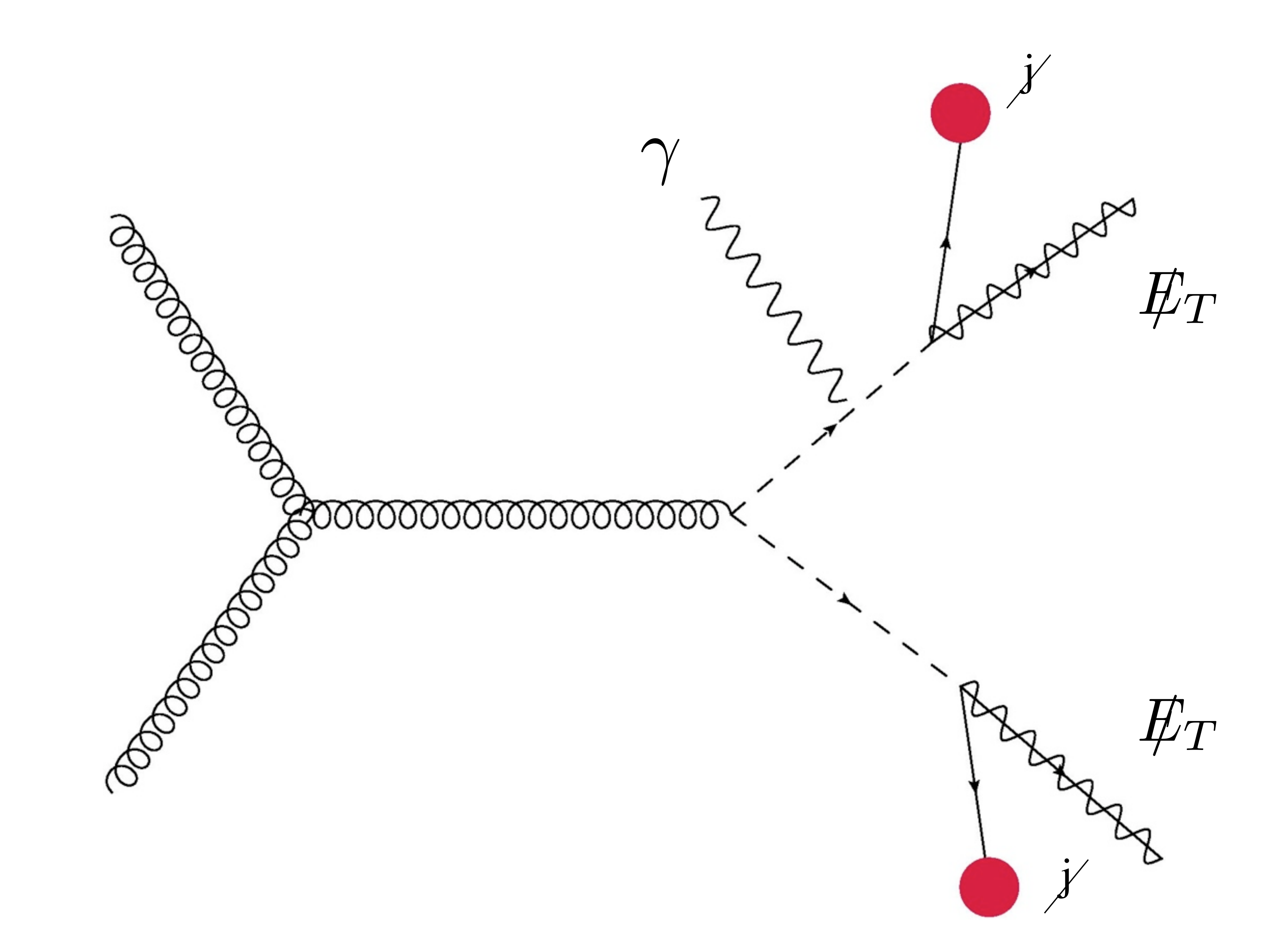}
\includegraphics[scale=0.2]{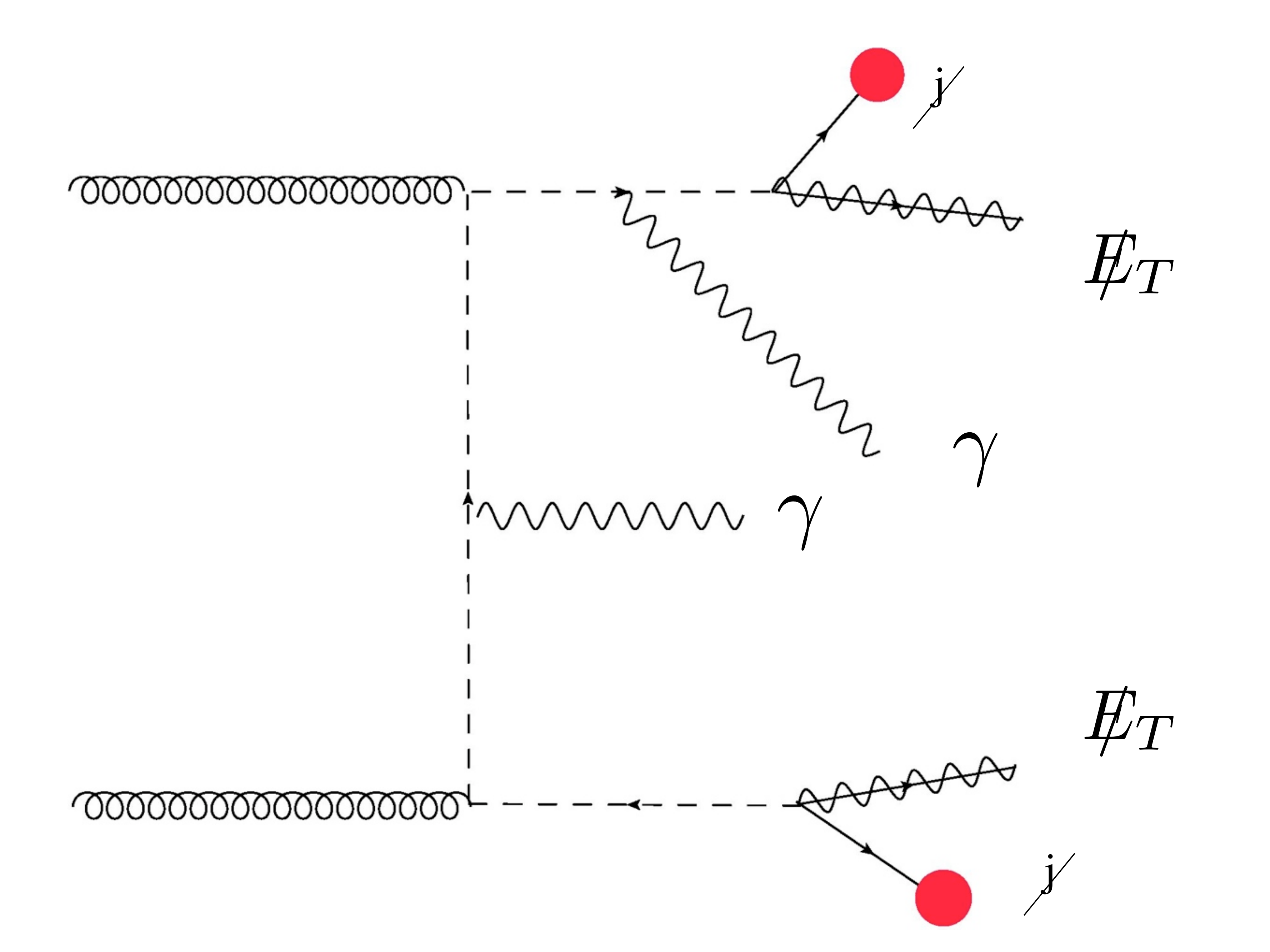}
\caption{Squark production and decay diagrams.}
\label{diagram}
\end{figure}

We restrict the squark production to the diagrams depicted in Fig.~\ref{diagram}), as these are model independent production mechanisms: they do not depend on the squark flavor, nor the gluino mass. 
Indeed, one could consider other production mechanisms with a t-channel exchange of a gluino, a diagram that is most relevant for first generation squarks. We will include this diagram in a second analysis where we make the additional assumption that all squarks are nearly degenerate.
 
As mentioned above, there are almost no bounds on squarks when the degeneracy with the neutralino is below roughly 10 GeV, 
only the   LEP limits extend to such small mass splittings.  
For sbottom quarks decaying exclusively into a b-quark and neutralino the LEP limit is then 100 GeV for a left-handed squark ~\cite{LEP_combined}.
The channel we examine here should therefore lead to a model independent limit on nearly degenerate squarks. 
 Note that it was suggested to search for stops nearly degenerate with neutralinos in missing energy plus one photon or jet at the Tevatron in ~\cite{Carena:2008mj}, as well as more recently at the LHC in ~\cite{Drees:2012dd}.

\section{The CMS analysis on monophoton and missing energy}

The search for events with a high-$p_T$ photon and missing energy with the CMS detector is presented in~\cite{CMSmonophoton}. The analysis uses 5.0 ${\rm fb}^{-1}$ of data collected in the 2011 run with $\sqrt{s}=7 \ {\rm TeV}$. The observed number of events is compatible with the standard model (SM) prediction and this result is interpreted in the analysis as an upper bound on WIMP-nucleon scattering cross section, when the $\gamma+\etm$ final state is interpreted as production of a dark matter particle in association with a high-$p_T$ photon. Also limits for models of large extra dimensions are given.

Here we will interpret the null result of this search as a model independent bound on squark production in the case of a compressed spectrum, i.e. when the mass splitting $\Delta m$ between the lightest squark and the LSP neutralino is small. As described above, this corner of SUSY parameter space is especially difficult to probe since the hard jets originating from the decays of squarks and gluinos are absent. In this case a squark pair decays practically invisibly into very soft jets and two LSPs, leaving no hard jets or leptons to use for triggering. To catch these events one can use initial or final state radiation of a hard photon or a jet, in which case the event is essentially a monophoton or a monojet accompanied with missing $E_T$. Here we will analyse the monophoton signal and leave the study of the monojet events for a future publication.

The most important cuts used in the event selection in the CMS analysis are
\bea
& & \textrm{ Photon quality (PHQ): } |\eta_{\ga}| < 1.44 \, \ , \,  p_T^{\ga} > 145 \gev \, \ \textrm{ and } \\
& &  \textrm{ Missing energy (MET): } \etm >  130 \gev \, .
\label{Eq:Cuts}
\eea
In addition, to reduce missidentification of leptons or pions as photons, several requirements for the quality of the photon candidate are in place as described in detail in~\cite{CMSmonophoton}. Events containing a jet with $p_T^j > 40 \ {\rm GeV}$ and $|\eta_j|<3.0$ within a cone of $\Delta R < 0.5$ around the photon axis are vetoed, as are any events that contain a track with $p_T > 20 \ {\rm GeV}$ that is $\Delta R > 0.04$ away from the photon candidate.

After those cuts there are 75 candidate events, whereas the estimated backgrounds are $75.1 \pm 9.4$.

 \section{The squark signal}

The CMS search is based on low-level objects, photons and missing energy, and whether the study is done at parton level or using a more sophisticated detector simulator is mostly relevant for reproducing the behavior of the photon isolation cut. Showering, and clustering of jets reduce the efficiency on passing PHQ cuts when the jets come too close to the photon. Extra jets also contribute to the systematic uncertainty in MET cuts, as mis-measurement of the jet momentum modifies MET determination.

In this section we describe the simulation tools, and the scenarios we consider: model-independent and degenerate scenario. Bounds on the degenerate scenario are much stronger than the model independent result, since they carry more assumptions on the squark spectrum.

 \subsection{Simulation details}
 In order to study the $\gamma+\etm$ signal associated with squark pair production we generate event samples with MadGraph5~\cite{Alwall:2011uj}. We use a model of MSSM with a hand made mass spectrum, where all supersymmetric particles except one squark mass eigenstate and one neutralino are heavy and decoupled, and the light squark and neutralino are nearly degenerate in mass. We also study a scenario where all first and second generation squarks are light and degenerate in mass, with and without a relatively light gluino.

The parton level events generated by MadGraph5 are passed to Pythia~\cite{Sjostrand:2006za} to simulate the effects of parton showering, and then to Delphes~\cite{Ovyn:2009tx} for a fast detector simulation. We use the default CMS-parameters for Delphes, and reconstruct jets with the anti-$k_T$ algorithm using $0.5$ for jet cone radius~\cite{CMSmonophoton}. As the event selection in this study depends on very clean signatures, i.e. an isolated high-$p_T$ photon and missing energy, the effects of the detector simulation compared to parton level events after Pythia turn out to be fairly small. For example the $p_T$-spectrum of the photon and the $\etm$-spectrum of the events are not affected much by the detector effects.  
 
 \subsection{The signal rate and systematics}\label{signalrate}

The signal we are looking at is
\bea
p \, p \to \slashed{E}_T + \gamma +X \ ,
\eea
where $X$ represents any object which does not overlap with the photon, and $\slashed{E}_T$ is the result of collider-stable neutral particles, e.g. a neutralino as the lightest SUSY particle in a scenario with R-parity. This final state, captured by the monophoton CMS search, can be interpreted in a very general way in SUSY scenarios. 

In this paper, our main concern is to provide a bound on squark masses as model-independent as possible. To do that we consider squark pair production in association with a photon coming from any leg, see Fig.~\ref{diagram}. The model independency comes from setting bounds on the squark mass under the following assumptions
\begin{enumerate}
  \item The squarks are pair produced through the diagrams depicted in Fig.~\ref{diagram}, i.e. the production is not assisted by any other SUSY particle (gluino, neutralino, $\ldots$).
  \item Only one type of squark mass eigenstate is produced (2 degrees of freedom). 
  \item The squark is nearly-degenerate in mass with the collider-stable, neutral particle, and hence the decay produces no reconstructed object. 
  \item No displaced vertex is reconstructed.
\end{enumerate}

To simulate these conditions, we consider a simplified model with one type of  squark $\tilde{q}$ and one neutralino $\tilde{\chi}^0$. We vary the mass separation $\Delta m=m_{\tilde{q}}-m_{\tilde{\chi}^0}$ from 1 to 30 GeV, as the monophoton search is competitive with the standard multijet+ MET~\cite{Aad:2011ib} only in that range of masses. 

In Sec.~\ref{results} we will also present results in the simplified model considered in the multi-jets+MET+X searches at ATLAS~\cite{ATLAS_CMSSM}, but in a range of masses the ATLAS search cannot probe. The simplified model consists of one gluino and degenerate first and second generation  LH and RH squarks. 
The searches in Ref.~\cite{ATLAS_CMSSM} do assume a fixed value for the neutralino mass, although most results do not change as long as $\Delta m \gg m_{\tilde{\chi}^0}$. Complementarily, in this paper we are going to explore the parameter space with small $\Delta m$.

Within these simplified models, the signal normalization is fixed at NLO by using Madgraph5~\cite{Alwall:2011uj} for the leading order production, and the effect of NLO using PROSPINOv2.1~\cite{Beenakker:1996ed}. We then present our results varying the central value by 50\%. 
 
  \begin{figure}[h!]
\centering
\includegraphics[scale=0.35]{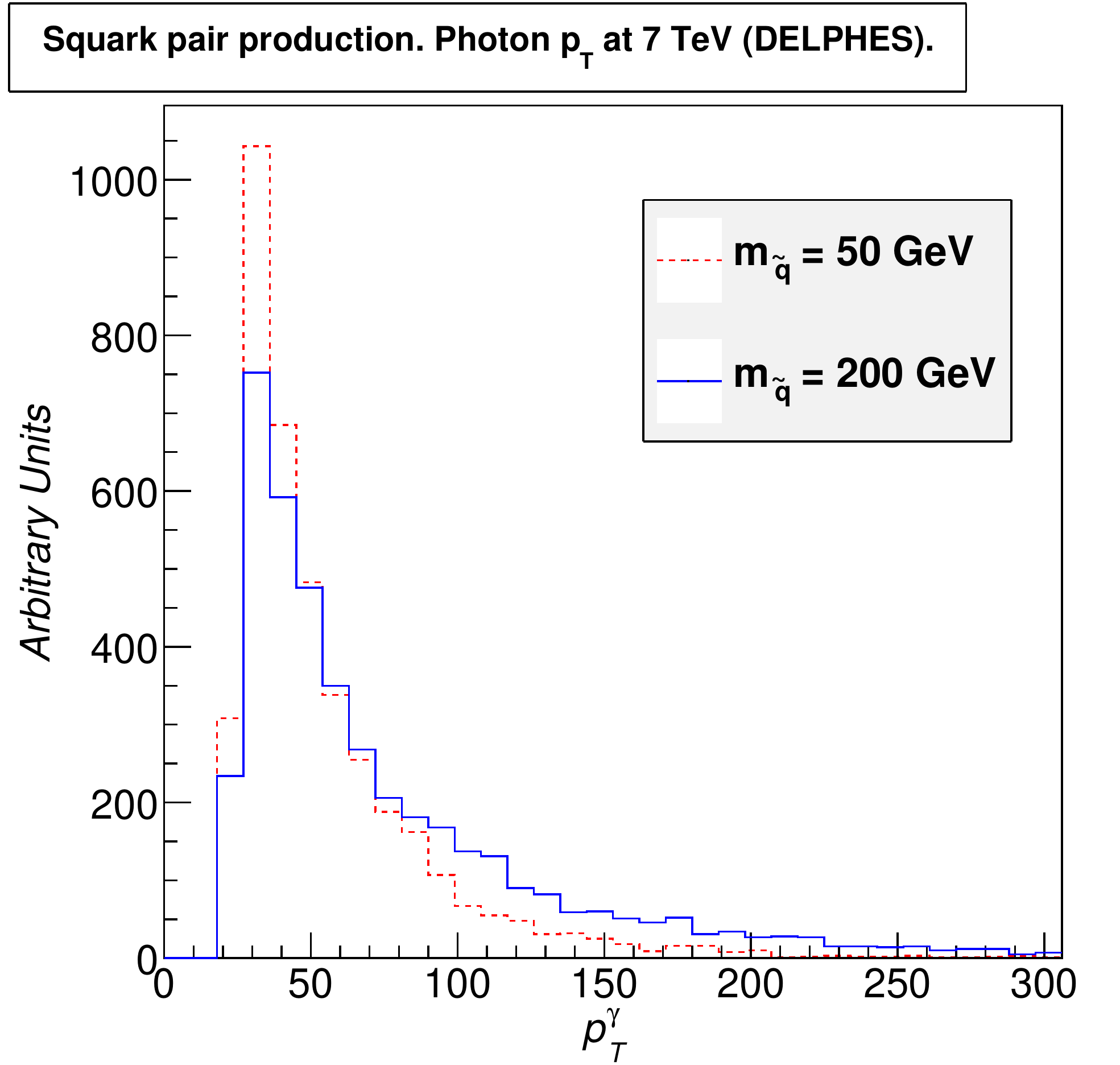}
\includegraphics[scale=0.35]{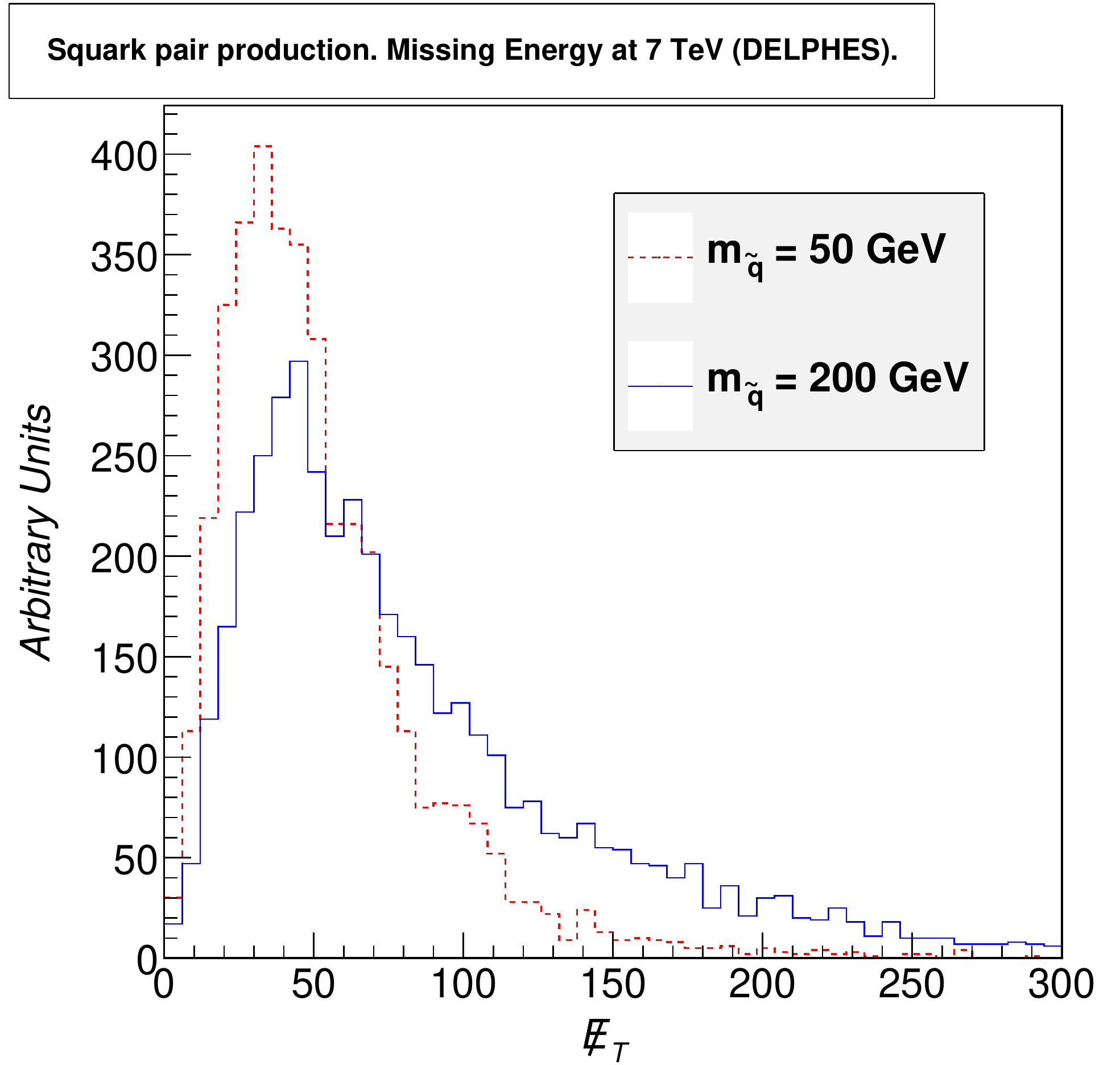}
\caption{Photon $p_T$ and missing energy distributions for different values of the squark/neutralino mass. Distributions are normalized to the same number of events.}
\label{Ptdistr}
\end{figure} 
 
The events coming from squark decays are central and not too boosted (well separated) in the low squark mass region. The photon momentum depends on the squark mass, as the kinematic configuration of initial quarks and gluons which leads to squarks of a given mass is different. Indeed, photons are more boosted as the squark mass increases, as the whole event has to be more energetic. This is shown in the left of Fig.~\ref{Ptdistr}, where we plot the photon $p_T$ distribution for two squark masses. Hard photons are on the tail of both distributions, but larger squark masses exhibit a slow increase on efficiency. The same behavior is obtained for the $\slashed{E}_T$ distribution changes. Since the neutral, stable particle mass is close in mass to the squark, the distribution is tilted as we increase the squark mass.

 In Table~\ref{cut7} we describe the efficiencies (in percent) of the PHQ and MET cuts for samples at 7 TeV. The MET cut number is {\it on top} of the PHQ cut. 

In this table, PHQ is mainly the efficiency of applying the cut $p_T^{\gamma}>$ 145 GeV, as we have generated events at parton level with $|\eta_{\gamma}|<$ 1.5. The numbers correspond to the situation of $\Delta m$=1 GeV (10 GeV). The variation of efficiencies in the range of $\Delta m$=1-30 GeV is $\lesssim 30$ \%, with the efficiencies decreasing as $\Delta m$ increases. When the squark and neutralino mass gap increases, the jets from the two-body decay, $\tilde{q}\to j \tilde{\chi}^0$, carry away more energy, leaving less to the radiated photon.

 These results show that the cut on photon momentum is the most stringent cut on the signal, as the MET cut is easily satisfied for $m_{\tilde{q}}\gtrsim$ 50 GeV. 
 
     \begin{table}
 \begin{center}
 \begin{tabular}{|c|c|c|}
 \hline
$m_{\sq}$ with $\Delta m$ = 1 (10) GeV  & PHQ   &  MET\\\hline
 50 & 2.6 (2.0) & 1.8 (1.14)  \\
 100 & 5.8 (4.2)  & 4.8 (3.3)  \\
  150 & 7.8 (7.8)  & 6.7  (6.7)  \\
   200 & 10.5 (9.7)  & 9.2 (8.5)  \\
 \hline
 \end{tabular}
 \end{center}
     \caption{Efficiencies of MET+PHQ cuts for LHC at 7  TeV for two mass splittings. }
\label{cut7}
 \end{table}


 \section{Bounds on squark masses}\label{results}
 
  Analyzing the simulated events, we determine the efficiency to the CMS cuts. Convoluting those with the signal normalization, one can draw a limit on the squark mass. This is the subject of this section, where we provide the interpretation of the CMS search in terms of a squark mass limit. We also estimate the reach with the 2012 data-set. 
  
 \subsection{Model-independent bound}

 We estimate the exclusion limit at 3$\sigma$ by
\bea
\sigma (m_{\tilde{q}}^{\textrm{ limit }}) = \frac{N_{obs}-N_{max,exp} (3 \sigma)}{{\cal L}} \simeq 5 \textrm{ fb} \, \ , 
\eea
where $N_{obs}$ is the number of observed events, $N_{max,exp}  (3 \sigma)$ is the maximum value of expected background events at 3$\sigma$. The lack of excesses at this level, imposes a bound on signal cross section times efficiency to the cuts, and this can be translated into a mass bound. 

\begin{figure}[h!]
\centering
\includegraphics[scale=0.3]{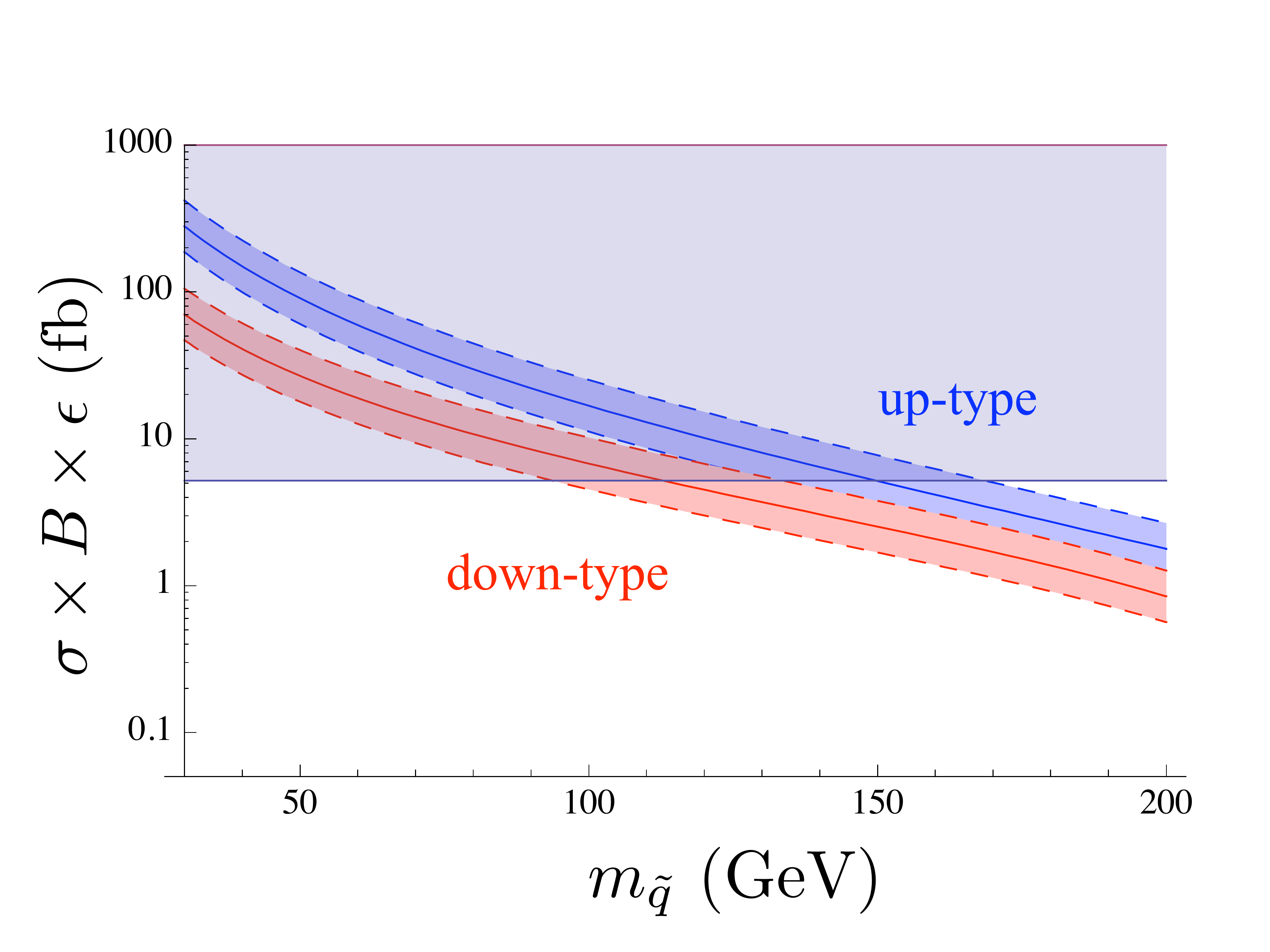}
\caption{Exclusion limits as a function of the squark mass (conservative). The blue (red) band corresponds to varying the PROSPINO central value by 50\%. The limit is on the production cross section times branching ratio to $\gamma$+MET times efficiency to the CMS cuts.}
\label{limit-MI}
\end{figure}

Under the assumptions described in Sec.~\ref{signalrate}, one can impose a bound on any type of squark. In Fig.~\ref{limit-MI}, the exclusion limits are shown as a function of the squark mass. As the cross section depends on the squark being up or down type, we plot in blue (red) the up (down) type squark limit. The  band corresponds to varying the PROSPINO central value by 50\%. The limit is on the production cross section times branching ratio to $\gamma$+MET times efficiency to the CMS cuts.

Using the results from the simulation depicted in Fig.~\ref{limit-MI}, one can impose a model independent bound valid when $\Delta m<10$~GeV
\bea
m_{\tilde{q}} \gtrsim 150 \, \,  (110) \textrm{ GeV, for {\it any} up (down) type squark,  }
\eea
 which goes beyond current bounds on a supersymmetric compressed spectra scenario.

\subsection{The limit for degenerate first and second generation squarks}

\begin{figure}[h!]
\centering
\includegraphics[scale=0.25]{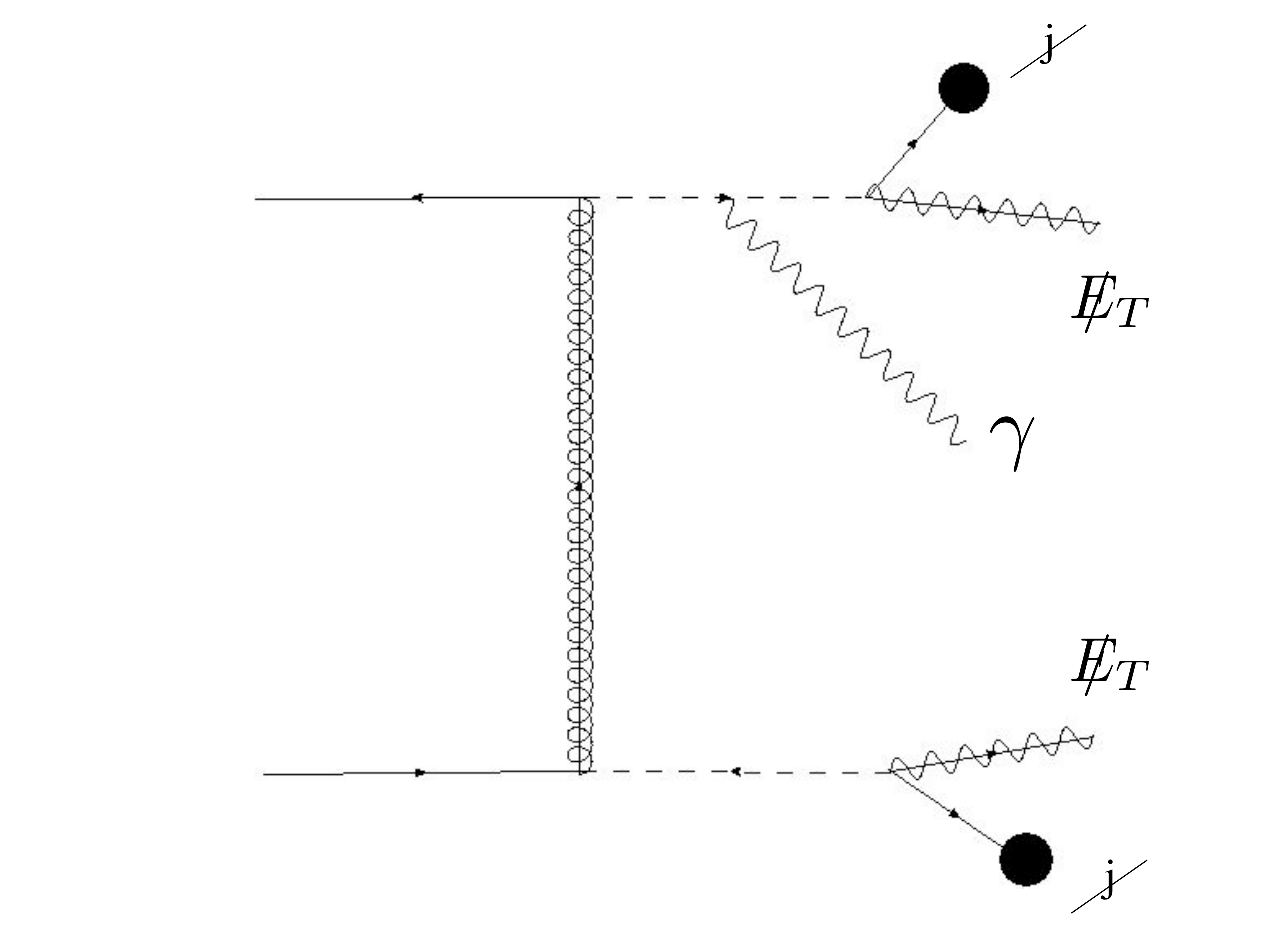}
\caption{Diagram of the gluino contribution to squark pair production.}
\label{tchannel}
\end{figure}

A more stringent bound can be obtained if further assumptions are made. To illustrate this, we obtain limits in the simplified model used in the ATLAS multijet+$\slashed{E}_T$ search. As we described in Sec.~\ref{signalrate}, this model assumes all first and second generation squarks are degenerate, an assumption inspired on solutions to the flavor problem in SUSY~\cite{Martin:2007gf}, but by no means general~\cite{Gedalia:2012pi}. The exclusion is then on a common squark mass, and also depends on the gluino mass, as the gluino contributes to the squark pair production. Indeed, for gluino masses $m_{\tilde{q}} \lesssim$ 1 TeV, the dominant production comes from the diagram in Fig.~\ref{tchannel}.

\begin{figure}[h!]
\centering
\includegraphics[scale=0.3]{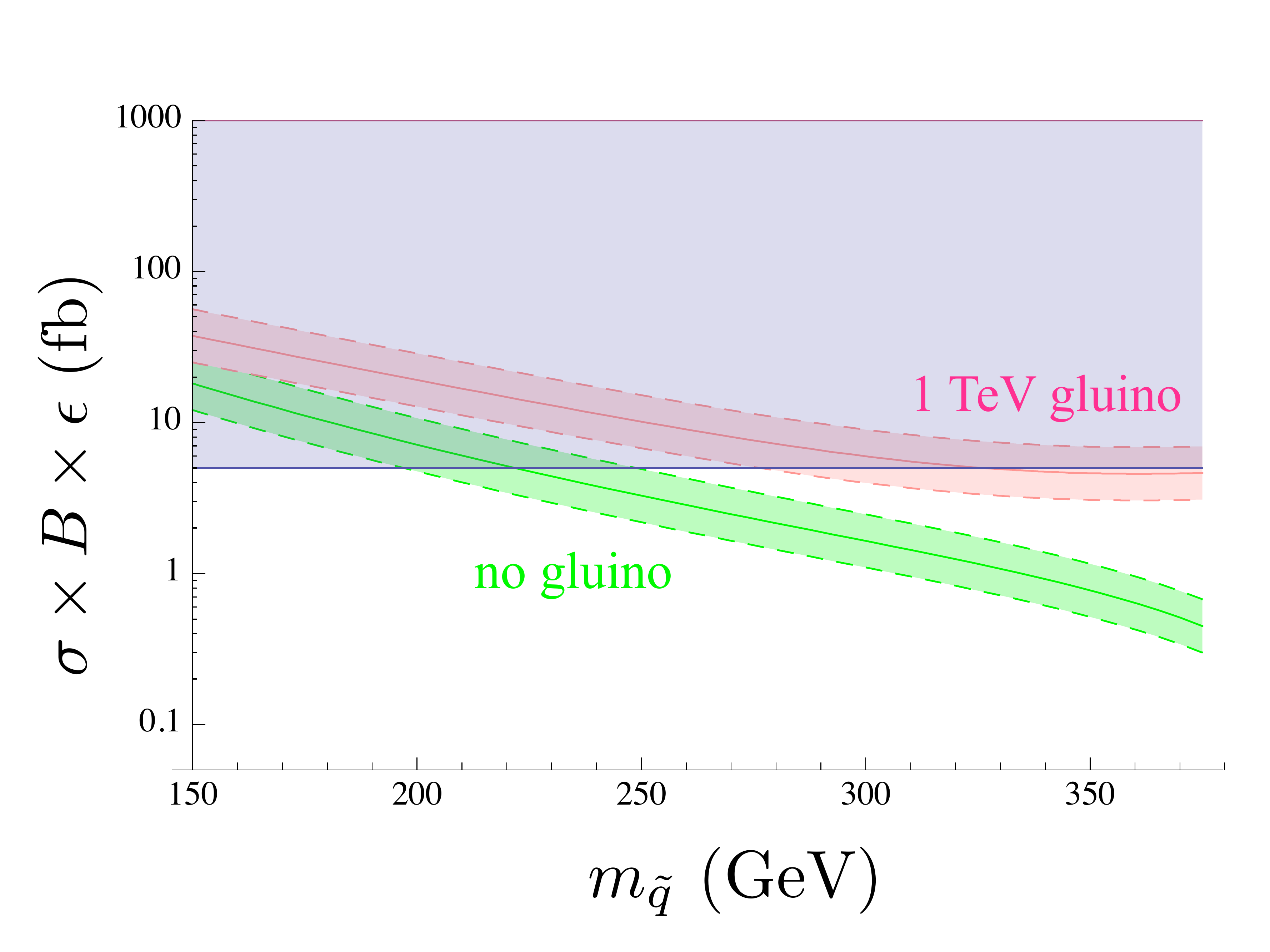}
\caption{Exclusion limits as a function of the common squark mass for a 1 TeV gluino, and in the case where the gluino is very heavy and does not participate in the squark cross section. The band corresponds to varying the PROSPINO central value by 50\%. The limit is on the production cross section times branching ratio to $\gamma$+MET times efficiency to the CMS cuts.}
\label{limit-MD}
\end{figure}

We obtain the signal normalization using PROSPINO at NLO, and translate those rates into a bound on the common squark mass, for a fixed value of the gluino mass~\footnote{Note that the rate depends on assuming that gluinos are Majorana particles, as in the MSSM. If one assumes that gluinos are Dirac particles, the rate would be smaller, and the exclusion less stringent~\cite{Heikinheimo:2011fk}.}. 
In Fig.~\ref{limit-MD}, we show those limits for a 1 TeV gluino and for a case with a very heavy gluino.

From Fig.~\ref{limit-MD}, one can extract a bound on the 1st and 2nd generation squark mass scale,
\bea
m_{\tilde{q}} \gtrsim   330 \, \,  (220) \textrm{ GeV, for } m_{\tilde{g}} = 1 \textrm{ TeV } (\infty).
\eea

Standard searches for SUSY  rely on missing energy in association with
hard objects from SUSY cascade decays. Searches with multijets and
missing energy are especially dramatic, but their strength quickly
disappears in a compressed spectrum scenario, the limits quoted above  are in this case the most stringent ones. 

To illustrate this point, we generated samples of squark pair
production, where the squark decays into jets and LSP. The efficiency
to the cuts on jet hardness is very low, and only the jets coming from
radiation (simulated by PYTHIA) is able to pass them. Therefore, the
interpretation of this search in terms of new physics depends
crucially on the understanding of the Monte Carlo used to simulate the
radiation jets.

To be specific, in a sample with $m_{\tilde{q}}=130$ GeV and $\Delta
m=$ 10 GeV, the efficiency in the bin of two or more jets (see the
selection cuts on this  bin in Ref.~\cite{Aad:2011ib}) is $5 \times
10^{-5}$, which leads to $\sigma\times B \times \epsilon$ of $\sim$ 1
fb, whereas the monophoton search leads to $\sigma\times B \times
\epsilon$ of order 10 fb.

\subsection{Reach in the 8 TeV run}

Finally, an update of the monophoton search in the 8 TeV run of the LHC would tighten the bounds on squarks if no signal is found. To estimate the improvement from this run, we will account for the change in cross section and efficiencies, and assume 20 fb$^{-1}$ of luminosity per experiment at the end of the 2012 run, which is a conservative assumption.

We are also going to assume CMS keep the same cuts as in the 7 TeV run. The PHQ cuts would scale up to 8 TeV, but we anticipate the MET cuts would need to be slightly raised. Nevertheless, we are going to estimate the reach using the same MET cut, as the higher values of $m_{\tilde{\chi}^0}$ we would be probing  in the 8 TeV run would anyway lead to a higher efficiency, even if the MET is raised. 

First, when moving from 7 to 8 TeV LHC, the increase of production cross section for the signal is a factor 1.3-1.6 in the range $m_{\tilde{q}}=100-400 $ GeV. Next, the efficiencies slightly increase as the cuts select high-$p_T$ objects, see Fig.~\ref{7to8}.
 
   
   \begin{table}[h!]
        \begin{center}
\begin{tabular}{|c|c|c|}
 \hline
$m_{\sq}$  & PHQ+MET (7 TeV) & PHQ+MET (8 TeV)   \\\hline
 100 & 5\% & 6\%  \\
 400 & 14\% & 16\% \\ \hline
 \end{tabular}
         \end{center}
         \caption{Estimated efficiencies for MET+PHQ cuts for LHC at 7 and 8 TeV.}
\label{7to8}
 \end{table}

 We can also study the effect of the most sizable background for the CMS search, the SM $Z\gamma$. By asking for a photon of $|\eta|<$1.44 and $p_T>$ 145 GeV, the effect of going from 7 to 8 TeV is 20\% increase in the cross section. 
 
 Therefore, the signal significance $S/\sqrt{B}$  would grow by a factor $\gtrsim$ 3, leading to an increase in the model independent bound on the squark mass of 20-30 GeV. This is a conservative estimate on the reach at 8 TeV, as we expect the analysis would improve further by tightening the MET cuts, and by the use of more sophisticated background rejection methods.

\section{Conclusion}

 Using the results of the CMS search for monophotons and missing energy at 7 TeV and ${\cal L}=4.7 fb^{-1}$, we have derived a lower bound on light squarks 
 when their mass splitting with the neutralino LSP is below the tenths of GeVs. 
 This channel allows to set conservative model independent limits  that extend past the existing collider constraints from LEP, Tevatron or the LHC. For any down-type squark the lower limit is set at 150 (110) GeV and for up (down)-type squarks. 
 Furthermore the bounds improve to 220 GeV if one assumes degenerate first and second generation squarks, 
 and to 330 GeV when including the  contribution of the gluino exchange diagrams with a 1 TeV gluino.
  A modest gain in sensitivity is expected from a naive extrapolation at 8 TeV. 
  
  Compared with monojet searches, high-$p_T$ photons are cleaner signatures, and their powerful background rejection compensates for the loss of production cross section. Moreover, the monophoton search scales-up in a straightforward manner at higher energies, providing a more robust limit than using monojet searches.

 These model independent limits can be applied to scenarios with compressed spectra that are otherwise difficult to probe at the LHC. They are therefore complementary to the standard searches relying on hard squark decay products, dominant for larger mass splittings. 
  They also provide  constraints on  the parameter space of the general MSSM, in particular  for models with a squark NLSP.
 Indeed scenarios with nearly degenerate squarks and neutralinos 
 can easily satisfy all constraints and in particular the relic density constraint since the important contribution of co-annihilation channels involving squarks reduce the relic abundance below the upper limit of WMAP.  
Although the bound derived when squarks of the first and second generation 
 are nearly degenerate with the LSP is more stringent, the dark matter sector can provide a competitive bound. Indeed  the light first generation squarks lead to a large cross section for direct detection that can be above the limit set by Xenon100.

 \providecommand{\href}[2]{#2}\begingroup\raggedright\endgroup

 \end{document}